# Electrically Conductive Diamond Membrane for Electrochemical Separation Processes

*Fang Gao\*, Christoph E. Nebel*

Fraunhofer Institute for Applied Solid State Physics (IAF), Tullastraße 72, Freiburg 79108, Germany

**\*Corresponding Author**

Email: fang.gao@iaf.fraunhofer.de

Telephone: +49(0)761-5159324

Fax: +49(0)761-515971324

**Abstract**

Electrochemically switchable selective membranes play an important role in selective filtration processes such as water desalination, industrial waste treatment and hemodialysis. Currently, membranes for these purposes need to be optimized in terms of electrical conductivity and stability against fouling and corrosion. In this paper, we report the fabrication of boron-doped diamond membrane by template diamond growth on quartz fiber filters. The morphology and quality of the diamond coating are characterized via SEM and Raman spectroscopy. The membrane is heavily boron doped (> $10^{21}$ cm$^{-3}$) with > 3 V potential window in aqueous electrolyte. By applying a membrane potential against the electrolyte, redox active species can be removed via flow-through electrolysis. Compared to planar diamond electrodes, the ~250 times surface enlargement provided by such a membrane ensures an effective removal of target chemicals from the input electrolyte. The high stability of diamond enables the membrane to not only work at high membrane bias but also to be self-cleaning via *in situ* electrochemical oxidation. Therefore, we believe that the diamond membrane presented in this paper will provide a solution to future selective filtration applications especially in extreme conditions.

**Introduction**

Porous membranes with tunable selectivity by external stimuli have been of great interest in chemical and biological separation processes.[1-7] In some of these instances, the external stimuli directly change the physiochemical nature of the membrane material, causing changes in permeability. Such mechanism applies to membranes which are pH-,[2-3, 7] temperature-,[6] photo-[5] and potential-sensitive.[1, 4] In some other cases when the membrane itself is electrically conductive, the ion permeability can be tuned by the membrane potential.[1] The local charge accumulation will change the trans-membrane movement of the ions. This concept has been proved by previous study on Au nanotubule membranes.[1] Because the amount of accumulated charge is proportional to the polarizable range, or potential window, of the certain type of membrane, the electrochemical window of the electrode material in aqueous electrolyte is very important for this application.

It has been proven that boron-doped diamond (BDD) a large potential window up to 3.5 V in aqueous electrolyte.[8] This value is much larger compared to traditional electrode materials such as gold or platinum. Moreover, it is known to be stable in a wide range of corrosive chemicals including fluoride,[9] alkaline[8] and a boiling mixture of $HNO_3$ and $H_2SO_4$.[9] Electrochemically, boron-doped diamond electrodes can be used in 2 M $H_2SO_4$ over a voltage range of -35 V to + 6 V,[10] or in a mixture of 1.0 M $HNO_3$ and 2.0 M NaCl at a current density of 0.5 A cm$^{-2}$ for up to 12 h with no evidence of damage.[11] These unique properties have been applied in electrochemical sensing and energy storage.[12-15] The idea of using electrically conductive diamond membrane for potential-controlled flow-through electrolysis has been mentioned in the early 2000s.[16] After that, although considerable progress in the fabrication of diamond membranes has been made,[14, 17-20] the report on the application is rare. In the ideal case, the control of ion movement is completely realized via electrostatic interaction.[1, 21] In the Debye-Hückel approximation, the potential changes exponentially with the distance from the electrode surface with the Debye length $\lambda$ as decay constant:[22]

$$\lambda = (\frac{2N_A I e^2}{\varepsilon \varepsilon_0 k_B T})^{-\frac{1}{2}} \tag{1}$$

, in which $\lambda$: Debye length, $N_A$: Avogadro constant, $e$: elementary charge, $I$: Ionic strength, $\varepsilon$: permittivity, $\varepsilon_0$: vacuum permittivity, $k_B$: Boltzmann constant, $T$: temperature. With this relation we can estimate that for a 25 mM phosphate buffer, the Debye length is less than 2 nm. In most cases for diamond membranes, the pore size is much larger than the Debye length: typically >100 nm,[16, 19] sometimes up to ~10 μm.[23-24] Therefore, the electrical field in the pore is completely shielded by the double layer. As a result, to achieve molecular or ionic level filtration, the goal for the membrane fabrication would be to shrink the pore to a size which is comparable to the Debye length in typical electrolyte. The result closest to this goal was reported in 2011 on a diamond-like-carbon membrane.[25] With a sacrifice of the diamond quality, the pore size was decreased to ~30 nm, which is still more than one order of magnitude larger than typical values for Debye length in aqueous electrolyte. As a result, although proof of concept was observed, i.e. under positive bias, the movement of cation was hindered and *vice versa*, there is only a factor ~2 difference in the ion diffusion rate between the "on" and "off" states.

In the last two years, there has been considerable progress in the fabrication of porous diamond materials, including membranes.[14-15, 19-20] Amoung these developments, the coating technology of nanocrystalline diamond on filter paper provides an efficient way for the fabrication of diamond membrane. These membranes have been used in energy storage[14] and analytical chemistry.[19] Particularly, the work by Ruffinatto *et al* in 2014 has for the first time shown that diamond-coated glass fiber filter can be used in protein trapping through the interaction of the target molecule with the diamond surface. However, no evidence of potential-tuned property of such membrane has been shown. Meanwhile, the possibility of such BDD membrane to be used in electrochemically tuned filtration has been proposed for long time without a report realization.[14, 16, 19-20]

In this paper, we fabricated boron-doped diamond porous membrane with previously reported technology, and applied it for electrochemical separation processes. The membrane is based on coating a fibrous quartz network by highly boron-doped diamond (B-concentration > $10^{21}$ cm$^{-3}$). The surface area as well as potential window is characterized electrochemically. The high-conductivity is confirmed via impedance spectroscopy. As an electrically conductive membrane, the electron-transfer processes can take place at the membrane surface when appropriate potentials are applied. Our work differs from previous concept in a way that the species of interest are retained (or removed) via Faradic processes instead of electrostatic repulsion. The wide potential window and extreme chemical stability of diamond play an important role in this application, because 1) the membrane will not decay under the high applied potential needed for target decomposition, and 2) water splitting is minimized compared to the decomposition/deposition process of the chemicals of interest. As an exemplar system, a mixture of $Cu^{2+}$, $Ni^{2+}$ and Rhodamine B was applied. By varying the applied potential between +1.5 and -1.5 V vs. Ag/AgCl (1 M KCl), the three substances can be selectively removed in the filtration process. After long-term use, the contaminated diamond membrane can be easily revived via an *in situ* oxidation process. Compared to other electrically conductive membranes, this product is metal-free, highly conductive and applicable in extreme conditions such as high temperature, highly corrosive conditions. Also, the substances which can be separated by this membrane are not limited to charged species. In general, the diamond membrane that we present here stands for a new category of electrically conductive membranes for versatile selective filtering purposes in the future.

**Experimental**

The diamond deposition is realized via microwave plasma chemical vapor deposition (MWCVD). Prior to the growth, the quartz filters used as growth templates (2-inch, Macherey-Nagel, Germany) are immersed in H-terminated nanodiamond seeding solution.[26] As is described previously, the negative dipole on $SiO_2$ surface will interact with the positively charged H-terminated nanodiamonds,[27] and a seeding layer of nanodiamond will form on the quartz fiber. For the MWCVD, unlike the other work which concentrates on the low or ultralow temperature diamond growth, we would like to grow high quality diamond with good boron-incorporation. Therefore, the growth conditions which are close to the normal diamond growth conditions (650 – 900 °C) were used (**table 1**).[28] After growth, the filters were cleaned in boiled mixture of $H_2SO_4$ and $HNO_3$ at 250 °C to remove the $sp^2$ content. For characterization purpose, the $SiO_2$ substrate can be removed by hydrofluoric acid to generate free-standing diamond paper.

**Table 1.** MWCVD growth parameters for diamond-coated quartz filter.

| Parameter | Values |
| --- | --- |
| Methane (% in $H_2$ mixture) | 1 |
| Microwave Power (W) | 2500 |
| Gas Pressure (mbar) | 30 |
| Temperature (°C) | 800±40 |
| [B]/[C] ratio | 1% |

The morphology of the membrane is characterized by scanning electron microscopy (SEM, Hitachi 4500, Japan). The quality of diamond in terms of phase purity and boron incorporation is characterized by Raman spectroscopy. The electrochemical properties are measured on a potentialstat (VMP3, BioLogic, France). Basic electrochemical measurements (cyclic voltammetry and impedance spectroscopy) are performed in a three-electrode set-up with a Pt wire counter electrode, an Ag/AgCl (1 M KCl) reference electrode. All the potential values mentioned in this work will be based on this reference unless otherwise stated. A BDD membrane with a diameter of 1 cm is the working electrode, and it lies on a stainless 316 foil (Goodfellow, Germany) which acts as a current collector.

The potential-controlled membrane permeability is investigated in a two-chamber set-up (**figure 1**). On the donor chamber, two identical membranes with a diameter of 1 cm are placed symmetrically. The membrane located between the two chambers is the working electrode; the other membrane is the counter electrode. In the center of the donor chamber, an Ag/AgCl reference electrode is placed to monitor the membrane potential. Initially, the donor chamber is filled with the electrolyte to be filtered (and constantly refilled to keep the same liquid level during the whole experiment). Due to the pressure difference, the liquid will flow through the membrane and into the acceptor chamber. A flow rate of ~0.2 mL/min was measured. With a potential on the membrane, the species which can enter the acceptor chamber can be controlled. $Ni^{2+}$, $Cu^{2+}$ and Rhodamine B (RB) are chosen as the probe species. The composition is 1 mM $NiSO_4$, 0.1 mM $CuSO_4$ and 1 μM Rhodamine B. 10 mM $H_2SO_4$ is added to prevent the hydrolysis of metal ions. 0.3 M $NaClO_4$ is used as supporting electrolyte. $Ni^{2+}$ and Rhodamine B are detected via high performance liquid chromatography (HPLC, Shimadzu, Japan) using a 1:1 isocratic flow of 20 mM pH 2.9 phosphate buffer and acetonitrile. $Ni^{2+}$ concentration is monitored by 392 nm adsorption with a retention time of 2.47 min; RB is monitored by 556 nm absorption at a retention time of 14.68 min. $Cu^{2+}$ is detected via stripping voltammetry. The liquid within is mixed 1:1 with 0.1 M $H_2SO_4$. A BDD electrode is applied as the working electrode. The accumulation condition is -0.5 V, 15 s. The square-wave voltammetry (SWV) is performed with a pulse height of 50 mV, a pulse width of 50 ms and a step height of 10 mV. The height of the copper stripping peak at -0.1 V is used to determine the copper concentration. The membrane self-cleaning test is done in the same set-up. 2.0 V is applied to the membrane *in situ*. The duration is 5 min.

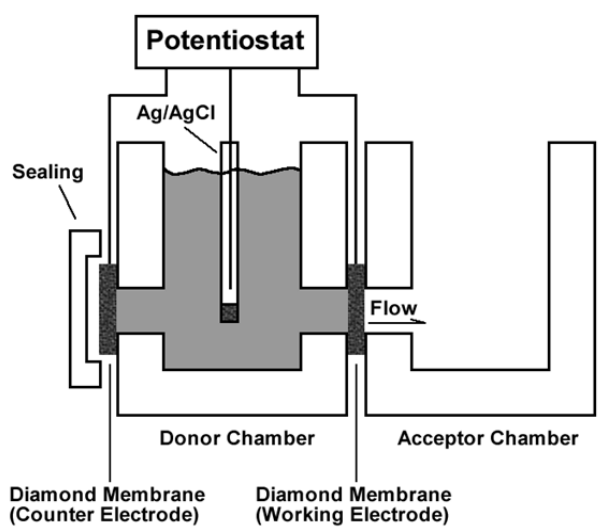

**Figure 1.** Schematic illustration of the two-compartment setup for diamond membrane mounting and testing.

**Results and Discussion**

The SEM images of the diamond membrane after acid cleaning are shown in **figure 2**. The top-view image (**figure 2a**) shows that the coating process does not change the fibrous nature of the filter paper, and the diamond-coated quartz fibers remain densely packed. The cross-section image (**figure 2b**) shows a coating reaches a depth of ~ 80 μm as is confirmed via zoom-in images shown in **figure 2(c) – (g)**. The thickness of BDD coated decreased with the distance from the surface of the template as a result of diffusion controlled growth. On the upper surface, the coating is close to 2 μm, with clear crystalline diamond grains. At a depth of 80 μm, the coating thickness decreases to less than 30 nm and the nanocrystalline film become discontinuous, which shows the maximum depth for diamond growth. In the previous work reported by Ruffinatto et al,[19] the depth of diamond growth was shown to be much higher (300 μm). This is probably due to the fact that their filter template has larger pore size (0.7 and 2.2 μm vs. 0.3 μm in our case). A thicker membrane helps the retention of target species and enhanced the stability. However, we can still remove the quartz fiber by hydrofluoric acid (HF) and pile several layers of BDD membrane up to reach a desired thickness.

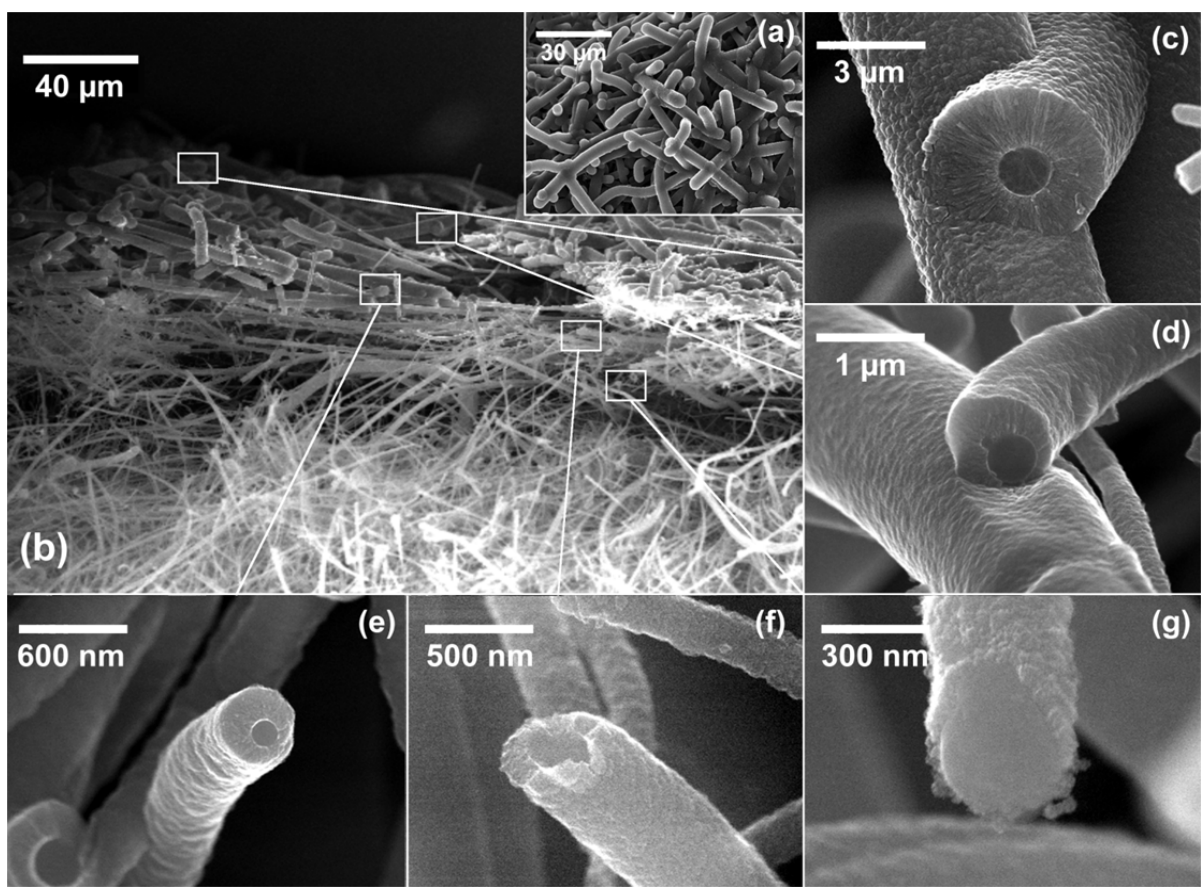

**Figure 2.** SEM images of top (a) and cross-section (b) view of the diamond-coated quartz fiber filter after acid cleaning; (c) – (g): zoom-in images at different depth in the cross-section showing the coating thickness.

The Raman scattering results on the top and bottom side of the diamond coating reveals more information about the diamond quality and the growth condition (**figure 3**). In order to perform Raman spectroscopy on both upper and lower side of the diamond coating, the quartz fiber was removed by HF. Another sample is measured directly after the coating process without any cleaning to show the non-diamond carbon deposited at the bottom part of the quartz paper. On the upper side, typical spectrum of highly boron-doped nanocrystalline diamond is clearly observed. The first order diamond peak appears at 1300 cm$^{-1}$. The large shift in the diamond peak together with two peaks at ~500 and ~1200 cm$^{-1}$ is a result of highly boron-doping. The non-diamond carbon (sp$^2$) impurities appeared as D and G peaks in the Raman spectra around 1350 (D) and 1520 – 1580 (G) cm$^{-1}$. If one compares the Spectra at the surface (spectrum I), inside (spectrum II) and at the bottom of the quartz filter (spectrum III), two trends can be found. One is the degrading diamond quality: the first order diamond peak become broader with increasing height for D and G peaks. At the bottom side of the filter paper, diamond peak actually disappears, and only D and G peaks are observed. This is explained previously by the inhibited etching effect of atomic hydrogen inside the template.[14] Therefore, an acid cleaning has to be applied to remove the sp$^2$-rich fraction in order to obtain a high quality diamond membrane. The other trend is the lower boron incorporation inside the template. It is well established that the incorporated boron in diamond can be estimated by the peak shift of the first order diamond peak.[29] If we compare spectra I and II in **figure 3**, the peak shift due to boron doping is much higher in spectrum I than II. If we calculate the boron concentration using the shift of first order diamond peak,[29] the doping level of the upper surface is 4×10$^{21}$ cm$^{-3}$, while at the bottom of the diamond coating the level is 1×10$^{21}$ cm$^{-3}$.

The reason for this difference is likely to be the temperature. Boron incorporation in CVD diamond is temperature dependent. Higher temperature in general helps the incorporation of B-dopants.[30] During the membrane growth, we control the substrate temperature via plasma power. Therefore, the upper surface which is directly exposed to the plasma will be more intensively heated. Because the filter paper used in this case poorly conducts heat, the temperature deep into the template will be lower than the surface. Although there is a difference in boron concentration on the top and bottom of the membrane, the doping level exceeds the requirement for metallic conductivity ($5\times10^{20}$ cm$^{-3}$) which enables the membrane to work as a porous electrode, as will be confirmed in electrochemical characterizations.

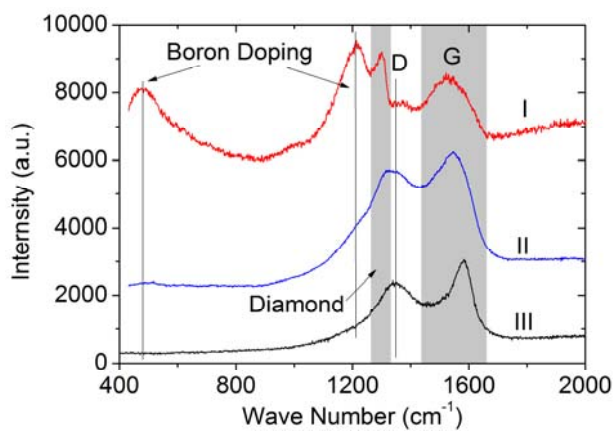

**Figure 3.** Raman spectroscopy at the top (I) and bottom (II) of the free-standing diamond membrane after acid cleaning and removal of quartz fiber template by HF; spectrum (III) shows the bottom of the quartz filter directly after coating without acid cleaning.

Compared to planar diamond electrodes, the surface enlargement is critical for an effective electrochemical removal of target substances. To obtain direct information about the surface enlargement, the capacitive current of the diamond membrane is measured via cyclic voltammetry. At a low scan rate, the capacitive current is proportional to the electrode surface area:[31]

$$I_c = C_n v A \qquad (2)$$

, where $C_n$ is the specific capacitance per unit area, $v$ is the scan rate, and $A$ is the active electrode surface area. The capacitive currents of a planar BDD and a BDD membrane are compared in **figure 4(a)**. Thanks to the templated-growth method, the diamond membrane inherited the large surface area provided by the quartz fiber filter. The surface enlargement is estimated to be ~250 times of a planar electrode. In our flow-through electrolysis, liquid that flows through will interact with this large surface area, and the trapping or removing of redox-active species will be more efficient than using planar electrodes.

One of the advantages of diamond in electrochemical applications is the wide potential window which is due to the slow kinetics of water splitting on the electrode.[8] As a result, the diamond membrane will provide a wide polarizable range in which chemical reactions can happen without or with limited interference with $H_2/O_2$ generation.[8] This property has been widely applied in electrochemical sensing of, for instance, $Cl_2$,[32] $H_2O_2$,[33] $O_2$,[34] and TNT.[35] However, the potential window is closely related to the quality of diamond, especially $sp^2$ content.[36-37] In this study, we compared the onset potential of $H_2/O_2$ evolution of the diamond membrane to a planar BDD electrode (**figure 4b**). The former has a potential window of 3.1 V (-1.6 – 1.5 V), which is only 0.5 V smaller than that of a high quality planar diamond electrode (-1.9 – 1.8 V). This wide potential window will in principal allow a wide range of applications of BDD membrane in aqueous electrolyte.

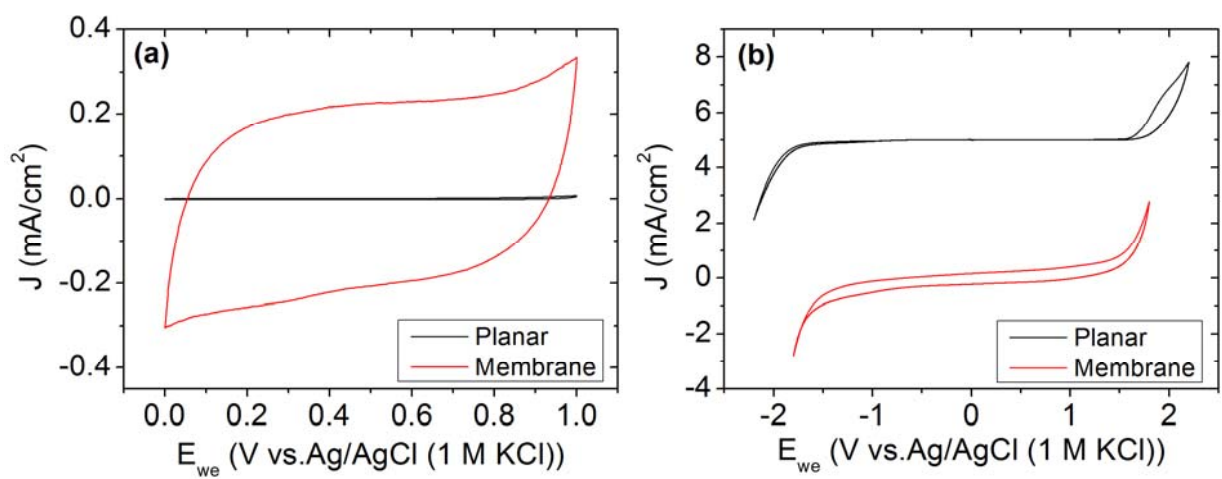

**Figure 4.** Electrochemical characterizations (scan rate: 100 mV/s for all cyclic voltammogram) of the diamond membrane electrode in comparison to a planar BDD electrode: (a) capacitive current, (b) potential window in 0.3 M $NaClO_4$

The ability of the porous membrane to trigger electrochemical reaction is confirmed using $[Fe(CN)_6]^{3-/4-}$ as a redox probe. Electron transfer processes on a semiconductor electrode are typically hindered by the depletion layer of charge carriers at the electrode side of the liquid-solid interface.[38] With a wide depletion layer which is a result of low doping, tunneling rate of electrons through it will be limited. Therefore, a high doping is necessary for efficient electron transfer at a semiconductor electrode. Figure 5 (a) shows a series of cyclic voltammogram of such membrane in 1 mM $K_3[Fe(CN)_6]$ in 0.1 M KCl. At slower scan rates, the peak separation ($\Delta E_p$) is as low as 63 mV which is comparable to the Nernst limit of 59.1 mV.[39] However, due to the known hindering effect of an oxidized BDD surface,[40] the redox reaction shows a quasi-reversible behavior with higher peak separations at higher scan rates (Figure 5b).[41] Meanwhile, the current of the cathodic and anodic peak follows the ideal linear relationship vs. the square root of scan rate. This indicates that both the oxidation and reduction reaction rate is controlled by semi-infinite linear diffusion of $[Fe(CN)_6]^{3-/4-}$ to the electrode surface.

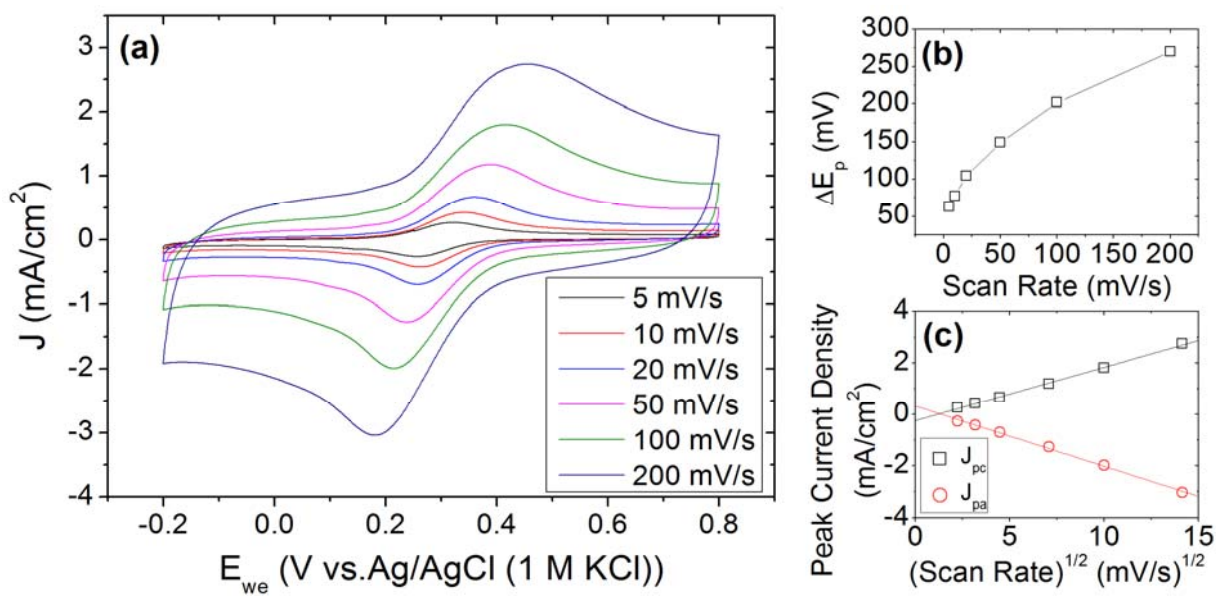

**Figure 5.** (a) Cyclic voltammogram of 1 mM $K_3[Fe(CN)_6]$ in 0.1 M KCl on BDD membrane with scan rates of 5 – 200 mV/s; (b) Anodic and cathodic peak separation vs. scan rate and (c) anodic and cathodic peak current density vs. the square root of scan rate.

To sum up, the basic characterization shows that the diamond membrane fabricated has a high phase purity, large surface area, wide potential window together with a highly active surface for electron transfer. Therefore, with an applied potential, this membrane is suitable for electrochemical treatment of liquid electrolyte which flows through it. Instead of electrostatic repulsion, we directly remove the target substance from the electrolyte via Faradic processes including deposition or decomposition. In this way, the mismatch between the pore size and double layer thickness is circumvented. There are many exemplar applications, such as waste water treatment (heavy metal ion precipitation), anti-fouling filter for environmental monitoring sensors, and analytical chemistry (interference removal). In the following section, we will show the selective removal of three kinds of ions by different applied membrane potentials.

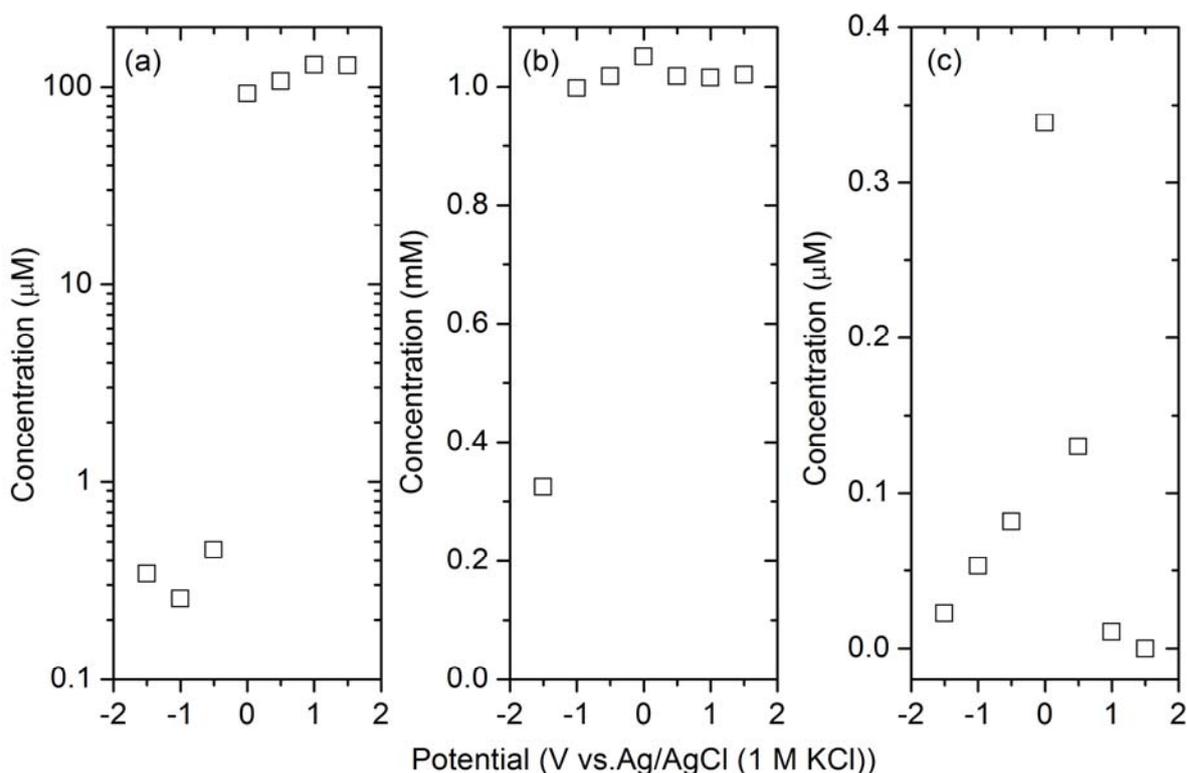

**Figure 6.** Concentration of (a) $Cu^{2+}$, (b) $Ni^{2+}$ and (c) Rhodamine B vs. membrane potential after filtration.

In our filtration experiment, $Cu^{2+}$, $Ni^{2+}$, and RB ions are chosen as the indicators because they all have well defined redox activities and are easy to detect. $Cu^{2+}$ will deposit as metal on the electrode at slightly negative potentials;[42] $Ni^{2+}$ will also deposit, but only at very negative potentials (< -1 V vs. Ag/AgCl);[43-44] RB are known to be decomposable on BDD electrodes at both negative and positive potentials.[45-47] The result of the electrochemically controlled filtration is shown in **figure 6**. At positive potentials (1.0 and 1.5 V), while the metal ions remain unaffected, RB is effectively removed (**figure 6c**). The concentration decreased to < 1% of the starting concentration. At the neutral potential, all the substance can go through the membrane without decomposing or deposition. At negative potentials (< -0.5 V), copper ions are reduced to <1% of the starting concentration (**figure 6a**), and also the dye is gradually removed (**figure 6c**). $Ni^{2+}$ is not affected until a very negative potential of -1.5 V due to the low reduction potential (**figure 6b**). Therefore, the electro-selectivity of the BDD membrane is clearly proven.

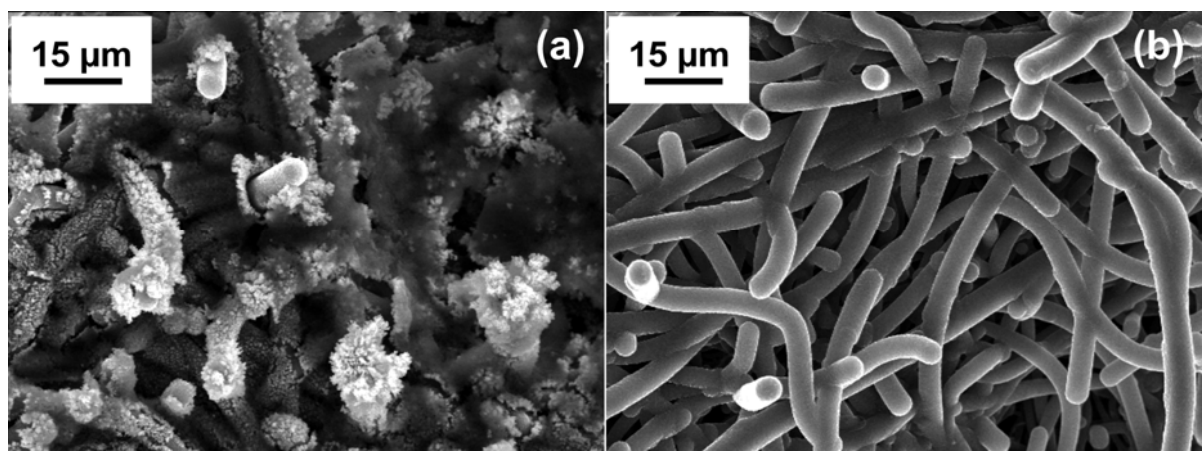

**Figure 7.** SEM images showing (a) membrane fouling after 1 h filtration of the given electrolyte mixture and (b) membrane after 5 min *in situ* regeneration with 2.0 V bias.

In our concept of filtration, the target is removed by electrolysis or electrodeposition. Therefore, the fouling of the membrane is unavoidable. However, thanks to the extreme stability of diamond, harsh electrochemical treatment can be applied to remove the contaminants. The in-situ self-cleaning properties has been reported widely previously on diamond-based sensors.[48] At very positive potentials (higher than 1.23 vs NHE), metallic residual will be converted to soluble ions in appropriate media and the generation of oxygen-related radicals effectively decompose organic compounds. This mechanism has been suggested previously for chemical oxygen demand (COD) determination electrochemically on a BDD electrode.[49-50] Extremely stable organic chemicals including ionic liquid were proved to be decomposable via oxidation processes on a BDD electrode.[51] In this study, similar method was applied to regenerate the active membrane surface. Strong oxidative potential effectively removes metallic deposits. Also, the oxygen-related radicals will help the removal of organic residues. **Figure 7(a)** shows the membrane after working with the above-mentioned electrolyte for 1 h at -1.5 V. As we can see, due to the large amount of deposits (Cu, Ni, and organic compounds from dye decomposition), the pores are heavily blocked, rendering the membrane unusable. After a cleaning step at 2.0 V *in situ*, the contaminants were completely removed, and the membrane was recovered to its original states (**figure 7b**).

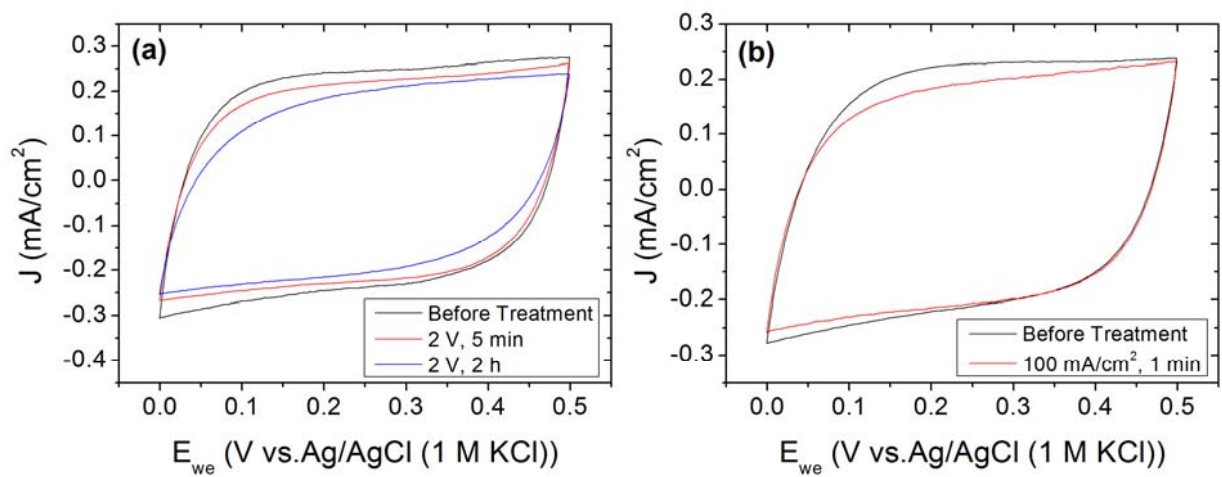

**Figure 8.** Cyclic voltammograms showing the capacitive current of the BDD membrane in 0.3 M NaClO$_4$ at 50 mV/s. (a) after 2 V biasing for 5 min and 2 h and (b) after 100 mA/cm$^2$ oxidation for 1 min.

To test the stability of the membrane against the *in situ* cleaning, the active surface area, represented by the capacitive background current, is measured before and after the high voltage (2.0 V) cleaning. **Figure 8 (a)** shows that the capacitive current drops from 0.241 mA/cm$^2$ to 0.223 mA/cm$^2$, indicating a 7.47% loss of active area in the first 5 min. This is probably due to the etching of graphitic grain boundaries or other non-diamond carbon species in the material shown in the Raman spectra.[52] If the high voltage treatment is prolonged, the further decay becomes slower: another 2 h holding at 2.0 V will reduce only another 7.88% drop in the capacitance current, which shows that the material is stabilized in the intensive oxidation process. In another more destructive test, a 100 mA/cm$^2$ current is driven through the membrane electrolyte interface, and the potential was raised to 4.5 to 5 V with strong bubbling from the membrane. Similar treatment was reported to be disastrous for sp$^2$ carbon-based materials.[53] However, for the diamond membrane, after 1 min of such treatment, the capacitive current drops only 7.45% (**figure 8b**). Please note that, in real application, the membrane needs to be clean only occasionally; the actually working potential was much lower than the potential applied in above-mentioned tests. Therefore, the life time of such membrane in the practical application will be much longer than the testing period. This high stability clearly distinguishes the diamond membrane from other $sp^2$ carbon materials like activated carbon. However, due to the considerably higher fabrication cost, our diamond membrane is not designed to replace activated carbon in all applications. Instead, it should find applications where corrosive environment is unavoidable, such as chlor-alkali process, or where the regular maintenance of the membrane is practically not possible such as deep-sea activities.

**Conclusions**

In this paper, we present a new application for porous conductive diamond membranes. Highly boron-doped diamond membranes have been fabricated and applied as an electrochemically switchable filter. The membrane reaches a thickness of ~80 μm with >$10^{21}$ $cm^{-3}$ boron-doping. It also shows large surface area, a wide potential window and fast electron transfer across the electrode surface. In the filtration test, it is shown that heavy metal ions can be trapped on the filter via electrochemical reduction, while organic compound can be decomposed via oxidation at positive potentials. Membrane fouling after long term use can be solved by a simple *in situ* oxidative regeneration process. Therefore, we believe that this type of diamond membrane will be suitable for applications such as selective removal of interfering chemicals for analytical instruments or anti-fouling filter for electrochemical sensor in environmental monitoring. With the possibility to further enlarge the filter to 6-inch, the membranes are also potentially useful for industrial applications such as waste water treatment, seawater desalination and the chlor-alkali process.

**Acknowledgement**

The authors thank Dr. René Hoffmann and Mr. Taro Yoshikawa for the helpful discussions on template diamond growth and Mr. Sascha Klingelmeier for the design of electrochemical reactor. We also thank Ms. Georgia Lewes-Malandrakis for the preparation of the diamond seeding solution. This project has received funding from the European Union's Horizon 2020 Program under Grant Agreement no. 665085 (DIACAT).